\documentclass[12pt]{article}

\usepackage{amssymb,amsfonts,amsmath}
\usepackage{longtable}
\usepackage{graphicx}
\usepackage{epsfig}
\usepackage{subfig}
\usepackage{url}

\begin{document}

\title{Quantifying replicability in systematic reviews: the $r$-value}

\maketitle
\begin{center}
Liat Shenhav \\
\emph{Department of Statistics and Operations Research, Tel-Aviv
university, Tel-Aviv, Israel. E-mail: liatshen@post.tau.ac.il}\\
Ruth Heller \\
\emph{Department of Statistics and Operations Research, Tel-Aviv
university, Tel-Aviv, Israel. E-mail: ruheller@post.tau.ac.il}\\
Yoav Benjamini \\
\emph{Department of Statistics and Operations Research, Tel-Aviv
university, Tel-Aviv, Israel. E-mail: ybenja@post.tau.ac.il}\\
\end{center}

\begin{abstract}
In order to assess the effect of a health care intervention, it is useful to look at an ensemble of relevant studies. The Cochrane Collaboration's admirable goal is to provide systematic reviews of all relevant clinical studies, in order to establish whether or not there is a conclusive evidence about a specific intervention. This is done mainly by conducting a meta-analysis: a statistical synthesis of results from a series of systematically collected studies.
Health practitioners often interpret a significant meta-analysis summary effect as a statement that the treatment effect is consistent across a series of studies. However, the meta-analysis significance may be  driven by an  effect in only one of the studies. Indeed, in an analysis of two domains of Cochrane reviews we show that in a non-negligible fraction of reviews, the removal of a single study from the meta-analysis of primary endpoints makes the conclusion non-significant.    
%, hence it is robust across the kinds of populations sampled. 
%The predicament is that in a meta-analysis, the summary effect significance can rely critically on a single study. 
Therefore, reporting the evidence towards replicability of the effect across studies in addition to the significant meta-analysis summary effect will provide credibility to the interpretation that the effect was replicated across studies.
We suggest an objective, easily computed quantity, we term the $r$-value, that quantifies the extent of this reliance on single studies. We suggest adding the $r$-values to the main results and to the forest plots of systematic reviews.  

\end{abstract}
%IN DISCUSSION ADD We feel that the Cochrane Review inititative is so important that even our minor contributions are important to improve its conclustions are important.

%BMJ REPRLICABILITY REFERENCE
%With the increasing concern about replicability over all branches of science, and in medicine as well REF BMJ, 

\vspace{0.5cm}

\section{Introduction}
In systematic reviews, several studies that examine the same questions are analyzed together. Viewing all the information is extremely valuable for practitioners in the health sciences. A notable example is the Cochrane systematic reviews on the effects of healthcare interventions. 
The process of preparing and maintaining Cochrane systematic reviews is described in detail in their manual \cite{Higgins11}. The reviews attempt to assemble all the evidence that is relevant to a specific healthcare intervention.

Deriving conclusions about the overall health benefits or harms from an ensemble of studies can be difficult, since the studies are never exactly the same and there is danger that these differences affect the  inference. For example, factors that are particular to the study, such as the specific cohorts in the study that are from specific populations exposed to specific environments, the specific experimental protocol used in the study, the specific care givers in the study, etc., may have an impact on the treatment effect. 

A desired property of a systematic review is that the
effect has been observed in more than one study, i.e., the overall
conclusion is not entirely driven by a single study. If a
significant meta-analysis finding becomes non-significant by leaving
out one of the studies, this is worrisome for two reasons: first,
the  finding may be too particular to the single study
(e.g., the specific age group in the study);  second, there is greater danger that the
significant meta-analysis finding is due to bias in the single
study (e.g., due to improper randomization or blindness). We view this problem as a replicability problem: the
conclusion about the significance of the effect is completely driven
by a single study, and thus we cannot rule out the possibility that
the effect is particular to the single study, i.e., that the effect
was not replicated across studies.

A replicability claim is not merely a vague description. A precise computation of the extent of replicability is possible. An objective way to quantify the evidence
that the meta-analytic findings do not rely on single studies is as follows. 
For a meta-analysis of several studies (N studies), the minimal replicability claim is that results have been replicated in at least two studies. This claim can be asserted if the meta-analysis results remains significant after dropping (leaving-out) any single study. We suggest accompanying the review with a quantity we term the $r$-value, which quantifies the evidence towards replicability of the effects across studies. 
The $r$-value is the largest of these $N$ meta-analysis $p$-values. Like a $p$-value, which quantifies the evidence against the null hypothesis of no effect, the $r$-value
quantifies the evidence against no replicability of effects. The
smaller the $r$-value, the greater the evidence that the conclusion
about a primary outcome is not driven by a single study. 
%In Section XXX

The report of the $r$-value is valuable for meta-analyses of narrow scope as well as of broad scope. 
In Section 5.6 of the manual \cite{Higgins11} the scope of the review question is addressed.
If the scope is broad, then a review that  produced a single meta-analytic conclusion may be criticized for `mixing apples and oranges', particularly when good biologic or sociological evidence suggests that various formulations of an intervention behave very differently or that various definitions of the condition of interest are associated with markedly different effects of the intervention. The advantage of a broad scope is that it can give a comprehensive summary of evidence. The narrow scope is more manageable, but the evidence may be sparse, and findings may not be generalizable to other settings or populations. If the $r$-value is large (say above 0.05) for a meta-analyses with a narrow scope, this is worrisome since the scope has already been selected, and the large $r$-value indicates that an even stricter selection that removes one single additional study can change the significant conclusion. If the $r$-value is large  for a meta-analyses with a broad scope, this is worrisome since the reason for the significant finding may be the single ``orange" among the several (null) ``apples".

We examined the extent of the replicability problem in systematic reviews. 
We found that there
may be lack of replicability in a large proportion of studies.
In Section \ref{sec-lack of replicability},  we show
that out of the 21 reviews with a significant  meta-analysis result
on the most important outcomes of interest published on breast
cancer, 13 reviews were sensitive to leaving one study out of the
meta-analysis. The problem was less pronounced in the reviews
published on influenza, where 2 reviews were sensitive to leaving one
study out of the meta-analysis, out of 6 updated reviews with
significant primary outcomes.
%A meta-analysis cannot serve as evidence towards replicability, since one study can drive the entire conclusion, see Example 1 XXX. In two domains XXX we examined, we say that in XXX of the systematic reviews it was enough to remove one study to reverse the conclusion.

\cite{AnzuresCarbera10}  write that a useful sensitivity analysis is one in which the meta-analysis is repeated, each time omitting one of the studies. A plot of the results of these meta-analysis, called an `exclusion sensitivity plot' by \cite{Bax06}, will reveal any studies that have a particularly large influence on the results of the meta-analysis. In this work, we concur with this view, but recommend the most relevant single number of summary information of such a sensitivity analysis be added to the report of the main results, and to the forest plot, of the meta-analysis. The code for the computation of the $r$-values and sensitivity intervals is available from the first author upon request.
%In addition, extensions of this approach are discussed in section...

\section{The lack of replicability in systematic reviews}\label{sec-lack of replicability}
%Show how prevalent the problem is for fixed effects, and for random effects, and include how prevalent fixed/random effects analysis are in the domains examined XXX
We took all the updated reviews in two domains: breast cancer and
influenza. Our eligibility criteria were as follows: (a) the review
included forest plots; (b) at least one primary outcome was reported as 
significant at the .05 level, which is the default significant level used in Cochrane Reviews; (c) the meta-analysis of at least one of the primary outcomes was based on at least three studies and (d) there was no reporting in the review
of unreliable/biased primary outcomes or poor quality of available evidence. 
We consider as primary outcomes the outcomes that were defined as primary by the review
authors, and if none were defined we selected the most important
findings from the review summaries and treated the outcomes for
these findings as primary. We limit ourselves to meta-analyses that
include at least three studies , since this is the minimum number of
studies for which even if the single studies are not  significant
the meta-analysis may still be non-sensitive (i.e., that a meta
analysis based on every subset of two studies can have a
significant finding).

In Cochrane reviews, the meta-analyses are of two types: fixed effect and random effects.
Under the fixed effect model all studies in the meta-analysis are assumed to share a common (unknown) effect $\theta$. Since all studies share the same effect, it follows that the observed effect varies from one study to the next only because of the random error inherent in each study. The summary effect is the estimate of this common effect $\theta$. Under the random effects model the effects in the studies, $\theta_{i}$ , $i = 1,2,...,N$, are assumed to have been sampled from a distribution with mean $\tilde {\theta}$. Therefore, there are two sources of variance: the within-study error in estimating the effect in each study and the variance in the true effects across studies. The summary effect is the estimate of the effects distribution mean  $\tilde {\theta}$. For details on estimation of these effects and their confidence intervals, see \cite{Higgins11}.  In this Section our results are based on the computations of the meta-analysis $p$-values as suggested in \cite{Higgins11}, for both fixed and random effects meta-analyses.
%but see Section \ref{sec-ex} for an alternative suggestion for the random effect meta-analysis.

In the breast cancer domain 48 updated reviews were published by the Cochrane Breast Cancer Group in the Cochrane library, out of which we analyzed 21 updated reviews that met our eligibility criteria (14, 8 , 4 and 1 reviews was excluded due reasons  a, b, c and d respectively). Out of the 21 eligible reviews, 13 reviews were sensitive to leaving one study out in at least one primary outcome. Moreover, in 8 out of 13 reviews all the significant primary outcomes were sensitive. The prevalence of sensitive meta-analyses was similar among the fixed effect and random effect meta-analyses, see Table \ref{Tab-BreastCancer}.  Among the 15 fixed effect meta-analyses,  6 reviews where sensitive in all their primary outcomes,  2 reviews were sensitive in  66\% of the primary outcomes, 1 review was sensitive in  50\% of the primary outcomes, and 6 reviews were not sensitive in any of their primary outcomes.
Among the 7 Random effect meta-analyses, 3 reviews were sensitive in all their primary outcomes,  2 review were sensitive in 50\% of their primary outcomes, and 2 reviews were not sensitive in any of their primary outcomes.

\begin{table}[ht]
\caption{Table of results for the breast cancer domain. The review name (column 2);the type of meta-analysis (column 3); the number of significant primary outcomes (column 4); the number of outcomes with $r$-values at most (0.01,0.05,0.1) (columns 5,6,7); the actual r-values of the primary outcomes, arranged in increasing order (column 8).The smaller the $r$-value, the stronger the evidence towards replicability. The rows are arranged by order of increasing sensitivity; the last 8 rows are sensitive in all primary outcomes. }\label{Tab-BreastCancer}
\scriptsize
\centering
\begin{tabular}{llllllll}
  \hline
   & &  &  number of   & \multicolumn{3}{c}{number of outcomes}& \\
      & &  &  significant   & \multicolumn{3}{c}{non-sensitive at level}& \\
 & Review & Random/Fixed &  outcomes & 0.01 & 0.05 & 0.1 & $r$-values \\
  \hline
  1 & CD004421 & Fixed & 2 & 2 & 2 & 2 & (1.300e-10, 1.405e-07) \\
  2 & CD003372 & Fixed & 2 & 2 & 2 & 2 & (4.000e-14, 5.368e-05) \\
  3 & CD002943 & Fixed & 2 & 2 & 2 & 2 & (9.853e-09, 0.0012) \\
  4 & CD006242 & Random & 2 & 2 & 2 & 2 & (2.580e-09, 0.03549) \\
  5 & CD000563 & Fixed & 1 & 1 & 1 & 1 & 1.28e-11 \\
  6 & CD008941 & Fixed & 1 & 1 & 1 & 1 & 1.025e-04 \\
  7 & CD005001 & Random & 1 & 0 & 1 & 1 & 0.0335 \\
  8 & CD003370 & Fixed & 1 & 0 & 1 & 1 & 0.05341 \\
  9 & CD003367 & Fixed & 2 & 1 & 1 & 1 & (7.440e-07, 0.18) \\
 10 & CD005211 & Random & 4 & 1 & 2 & 2 & (0.0017, 0.0167, 0.1231, 0.178) \\
 11 & CD003474 & Random & 2 & 1 & 1 & 1 & (0.0463, 0.253) \\
 12 & CD003366 & Fixed & 3 & 1 & 1 & 1 & (3.200e-05, 0.21, 0.38) \\
 13 & CD003139 & Fixed & 3 & 1 & 1 & 1 & (0.1053, 0.1852, 0.002) \\
 14 & CD006823 & Random & 1 & 0 & 0 & 1 & 0.08 \\
 15 & CD004253 & Random & 1 & 0 & 0 & 0 & 0.1028 \\
 16 & CD005002 & Fixed & 1 & 0 & 0 & 0 & 0.15 \\
 17 & CD008792 & Fixed & 1 & 0 & 0 & 0 & 0.24 \\
 28 & CD007077 & Fixed & 1 & 0 & 0 & 0 & 0.3 \\
 19 & CD002747 & Fixed & 1 & 0 & 0 & 0 & 0.9641 \\
 20 & CD007913 & (Random,Fixed) & 2 & 0 & 0 & 0 & (0.0712,0.0756) \\
 21 & CD003142 & Fixed & 2 & 0 & 0 & 0 & (0.1243,0.1827) \\
   \hline
\end{tabular}
\end{table}

In the influenza domain 25 reviews were published by different groups (e.g., Cochrane Acute Respiratory Infections Group, Cochrane Childhood Cancer Group etc.) in the Cochrane library, out of which we analyzed 6 updated reviews that met our eligibility criteria  (9, 2 , 7 and 1 review was excluded due reasons  a, b, c and d respectively). Our results are summarized in Table \ref{Tab-Influenza}. Out of the 6 eligible reviews, 2 reviews were sensitive to leaving 1 study out. Among the two fixed effect meta-analyses reviews, one review was sensitive in all primary outcomes and one review was not sensitive in all primary outcomes.
Among the five reviews with random effect meta-analyses, 1 review was sensitive in 40\% of the primary outcomes, and four reviews were not sensitive in any of their outcomes.

\begin{table}[ht]
\caption{Table of results for the influenza domain. The review name (column 2);the type of meta-analysis (column 3); the number of significant primary outcomes (column 4); the number of outcomes with $r$-values at most (0.01,0.05,0.1) (columns 5,6,7); the actual r-values of the primary outcomes, arranged in increasing order (column 8).The smaller the $r$-value, the stronger the evidence towards replicability. The rows are arranged by order of increasing sensitivity. The value $0.0001^*$ indicates that the $r$-value was smaller than 0.0001.}\label{Tab-Influenza}
\scriptsize
\centering
\begin{tabular}{llllllll}
  \hline
   & &  &  number of   & \multicolumn{3}{c}{number of outcomes}& \\
      & &  &  significant   & \multicolumn{3}{c}{non-sensitive at level}& \\
      & Review & Random/Fixed &  outcomes & 0.01 & 0.05 & 0.1 & $r$-values \\
  \hline
  1 & CD001269 & (Fixed,Random) & 4 & 4 & 4 & 4 & ($0.0001^*$, $0.0001^*$,0.0014, 0.0188) \\
  2 & CD001169 & Random & 4 & 4 & 4 & 4 & ($0.0001^*$ ,0.0014 ,0.0016, 0.0025) \\
  3 & CD004879 & Random & 4 & 4 & 4 & 4 & ($0.0001^*$ ,$0.0001^*$ ,0.0007, 0.006) \\
  4 & CD002744 & Random & 1 & 1 & 1 & 1 & 0.0009 \\
  5 & CD008965 & Random & 5 & 1 & 3 & 3 & ($0.0001^*$, 0.0108, 0.0471 ,0.118, 0.1206) \\
  6 & CD005050 & Fixed & 1 & 0 & 0 & 0 & 0.9888 \\
   \hline
\end{tabular}
\end{table}

The influenza domain has a much smaller number of reviews with significant primary results than the breast cancer domain. In the influenza domain, most of the reviews have non-significant endpoints or low quality of evidence.

%\section{Notation}

%Let $\theta_i$ be the (unknown) treatment effect in study $i$.
%Fixed effect meta-analysis XXX
%Random effect meta-analysis XXX

%\section{The $r$-value}
%Let $N$ be the number of studies.
%A replicability claim is a claim that the conclusion remains significant (e.g., rejection of the null hypothesis of no treatment effect) using a meta-analysis of each of the $\binom{N}{u}$ subsets of $N-u+1$ studies, where $u=2,\ldots, N$ is a parameter chosen by the investigator. Specifically, for $u=2$, a replicability claim is a claim that the conclusion remains significant  using a meta-analysis of each of the $N$ subsets of $N-1$ studies.

%The $r$-value is the smallest significance level at which we claim replicability.
%For fixed effects, it is the p-value for the test of the null hypothesis that the treatment has no effect except possibly in u-1 studies. For $u=1$, it is the maximum of the $N$ $p$-values from the $N$ meta-analyses of $N-1$ studies, for the test of  the null hypotheses of no treatment effect. See Appendix XXX for detailed computations when the hypotheses are two sided.

%Claiming replicability whenever the $r$-value is at most $\alpha$ guarantees control over a false replicability claim at level $\alpha$, see \cite{conj} for a simple proof.

%\newpage
\section{Calculating and reporting the r-value: examples}\label{sec-ex}
In this section  we shall give examples %in both the breast cancer and influenza domains,
of sensitive and non-sensitive (fixed and random effect)
meta-analyses in the breast cancer domain. For examples in the influenza domain, see Appendix
\ref{app-influenza}. For each example, we shall compute the $r$-value, which is based on the $N$ leave-one out meta-analysis $p$-values, as well as the sensitivity interval, which is the union of these $N$ meta-analysis confidence intervals. The detailed computations are given in Appendix \ref{app-computations}. We shall show  how to incorporate these new quantities in the  Cochrane reviews' abstract and forest plots .

%We shall give examples ,
%of sensitive and non-sensitive (fixed and random effect)
%meta-analyses in the breast cancer domain. For examples in the influenza domain, see Appendix
%\ref{app-influenza}. For each example, we shall compute the $r$-value. The $r$-value is
%the smallest significance level at which we claim replicability. For
%fixed effects, it is the p-value for the test of the null hypothesis
%that the treatment has no effect except possibly in a single study.
%See Appendix \ref{app-computations} for the exact computations. Claiming replicability
%whenever the $r$-value is at most $\alpha$ guarantees control over a
%false replicability claim at level $\alpha$, see \cite{conj} for a
%simple proof.

\begin{figure}[ht]
\centering
\includegraphics[width=0.4\textwidth, height=0.3\textheight]{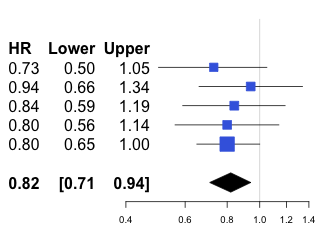}
\includegraphics[width=0.4\textwidth, height=0.3\textheight]{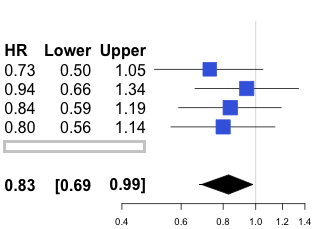}
\caption{The forest plot in Review CD006242 (Left) and excluding study 5 (Right). The $r$-value was  0.03549. The sensitivity interval, [0.685,0.987],  is the confidence interval excluding study 5 (the black diamond in the right panel) with an additional (very small) left tail. The axis is on the logarithmic scale.}\label{fig-Trastuzumab}.
\end{figure}

Our first example is based on a meta-analysis in review CD006242, analyzed by the authors as a random effect meta-analysis, which is
non-sensitive and thus has a small  $r$-value. The objective of review CD006242 
was to assess the efficacy of therapy with Trastuzumab  in
women with HER2-positive metastatic breast cancer. Only one of the
studies was (barely) significant, and the remaining four studies had
non-significant effects at the .05 significance level. However, when combined in a meta-analysis
the evidence was highly significant, and the review conclusion was
that Trastuzumab improved overall survival in HER2-positive women
with metastatic breast cancer, see the left panel of Figure
\ref{fig-Trastuzumab}. This is a nice example that shows how a
meta-analysis can increase power. Even after removing the
single significant study (study number 5) there was still a
significant effect in the meta-analysis at the 0.05 level; see the right panel of Figure
\ref{fig-Trastuzumab}. The $r$-value is 0.03549 based on the meta-analysis computations as suggested in \cite{Higgins11}. In a recent paper, \cite{IntHout14} suggested an alternative random effect meta-analysis, which controls the type I error rate more adequately. The $r$-value is 0.0366 based on the meta-analysis computations as suggested in \cite{IntHout14}.

\begin{figure}[ht]
\centering
\includegraphics[width=1\textwidth, height=0.5\textheight]{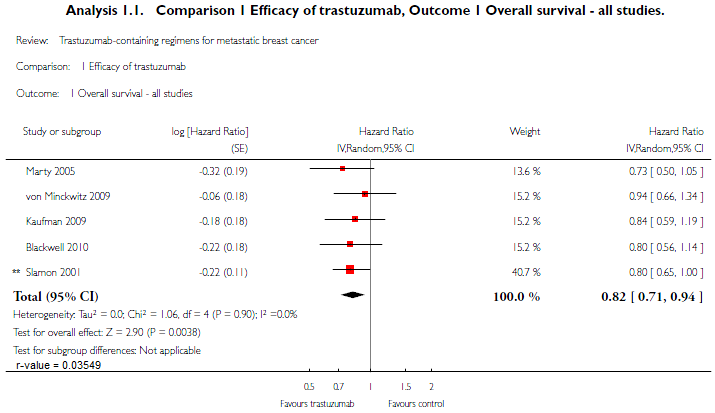}
\caption{The forest plot in the original Review CD006242 including the $r$-value, which  was  0.03549. The asterisks indicate which study was excluded for the $r$-value computation.}\label{fig-Trastuzumab-orig}
\end{figure}

We suggest accompanying the original forest
plot with this $r$-value, see Figure \ref{fig-Trastuzumab-orig}. The significant meta-analytic
conclusion can therefore be accompanied by a statement that the
replicability claim is established at the .05 level of significance.
This is a stronger scientific claim than that of the meta-analysis,
and it is supported by the data in this example. In the main results of Review CD006242 the authors write 
"The combined HRs for overall survival and progression-free survival favoured the trastuzumab-containing regimens (HR 0.82, 95\% confidence interval (CI) 0.71 to 0.94, P = 0.004; moderate-quality evidence)". To this, we suggest adding the following information ``This result was replicated in more than one study ($r$-value = 0.03549)". %0.685 is the smallest LB 

\begin{figure}[ht]
\centering
\includegraphics[width=0.4\textwidth, height=0.3\textheight]{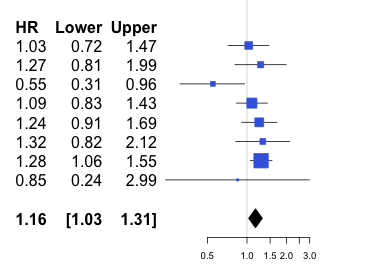}
\includegraphics[width=0.4\textwidth, height=0.3\textheight]{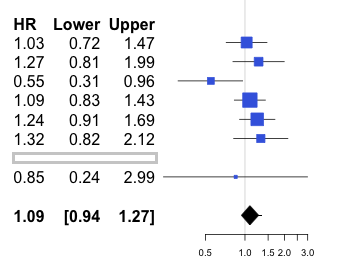}
\caption{The forest plot in Review CD008792 (Left) and excluding study 7 (Right). The $r$-value was  0.24. The sensitivity interval, [0.94, 1.356], is the confidence interval excluding study 7 (the black diamond in the right panel) with an additional (small) right tail. The axis is on the logarithmic scale.}\label{fig-Combination}
\end{figure}

Our second example, also from the breast cancer domain, is based on a meta-analysis in review CD008792, that was analyzed by the authors as a fixed effect meta-analysis. In this example the fixed effect meta-analysis was sensitive. The objective of 
Review CD008792 was to assess the effect of combination
chemotherapy compared to the same drugs given sequentially in women
with metastatic breast cancer. In the meta-analysis a significant
finding was discovered, see the left panel of Figure
\ref{fig-Combination}. However, note that the different studies seem
to have different effects. Nevertheless, the 
review conclusion was that the combination arm had a higher risk of
progression than the sequential arm. After removing study number 7,
there was no longer a significant effect in the meta-analysis, see
the right panel of Figure \ref{fig-Combination}. The $r$-value was
0.24.  The replicability claim was not established at the .05 level
of significance. This lack of replicability, quantified by the $r$-value,  cautions practitioners that the significant meta-analysis finding may depend critically on
a single study. 

\begin{figure}[ht]
\centering
\includegraphics[width=1\textwidth, height=0.5\textheight]{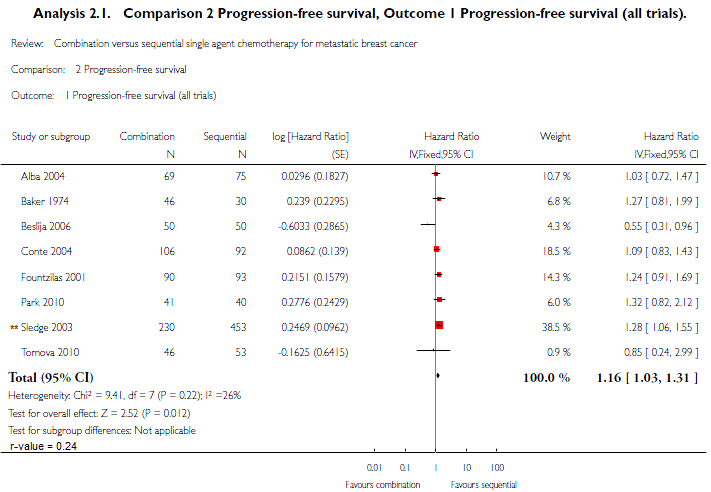}
\caption{The forest plot in the original Review CD008792 including the $r$-value, which  was  0.24. The asterisks indicate which study was excluded for the $r$-value computation.}\label{fig-Combination-orig}
\end{figure}

We suggest accompanying the original forest plot
with this $r$-value, see Figure \ref{fig-Combination-orig}.
In the main results of Review CD008792 the authors write  "The combination arm had a higher risk of progression than the sequential arm (HR 1.16; 95\% CI 1.03 to 1.31; P = 0.01) with no significant heterogeneity". To this, we suggest adding the following information ``We cannot rule out the possibility that this result is based on a single study ($r$-value = 0.24)". %1.356 is the largest UB
%Note: this result was not replicated in more than one study (r-value = 0.24).

In the right panels of Figures \ref{fig-Trastuzumab} and
\ref{fig-Combination}, the meta-analysis confidence intervals that
would have been computed had we considered only this specific subset of
studies is shown. The sensitivity intervals has an additional tail in the direction favoured by the data.

%\section{Efficient algorithm}
%For fixed effect: no need to go over all subsets since know from estimated effects, SE, and sample sizes which study should be taken out to lead to the most significant result XXX?

%Similarly for random effect: what is the efficient algorithm XXX?

%\section{Extensions}
%\subsection{Pointing out the replicated studies}
%For fixed effects, when concluding that the finding is replicated in at least $u$ studies, it may be of interest to identify the studies. For this purpose it is necessary to test for replicability using Holm/Simes/Min p-value ...

%\subsection{Different combining methods}
%For fixed effects, can combine the $p$-values instead of the fixed effect meta-analysis combining method, as detailed in \cite{conj}. SHOW AN EXAMPLE WHEN IT MATTERS?
%\subsection{Robust estimation of treatment effect (and confidence intervals)}
%Downward biased if all from the same famiy? Otherwise, if indeed u-1 are not have unbiasedness?

\section{Methodological extensions}
\subsection{A lower bound on the extent of replicability}
A review is less sensitive than another review if a larger fraction
of studies are excluded without reversing the significant
conclusions. We can calculate the meta-analysis significance not only after dropping each single study, but also after dropping all pairs of studies, triplets of studies etc. Each time we calculate the maximum $p$-value and stop at the first time it exceeded $\alpha$. The bigger the number of studies that can be dropped, the stronger the replicability claim.

%We can ask whether excluding at least $u-1$ studies will reverse the conclusion, with $u\in \{2,\ldots, N\}$. Testing in order at significance level $\alpha$ (i.e., for $u=2,3,\ldots$) using the $r$-value computed by taking the maximum over $\binom{N}{u-1}$ subsets of $N-u+1$ studies, results in a $1-\alpha$ confidence lower bound on the number of studies with an effect in a fixed-effect meta-analysis (see \cite{Heller11} for a proof). 

\begin{figure}[ht]
\centering
\includegraphics[width=1\textwidth, height=0.5\textheight]{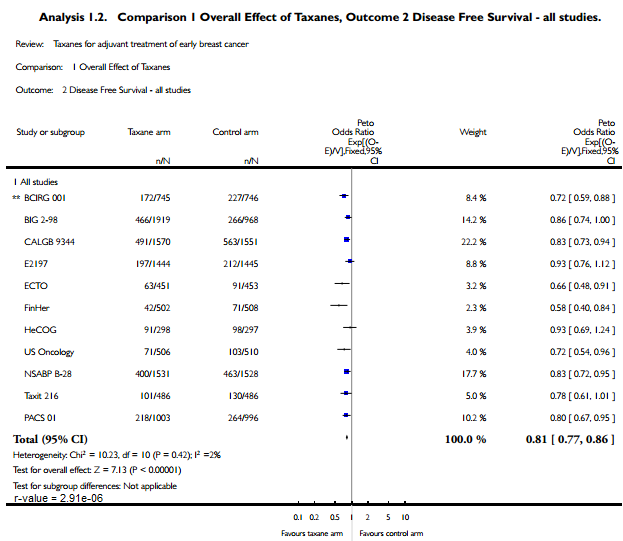}
\caption{The forest plot in the original Review CD004421 including the $r$-value, which  was  2.91e-06. The asterisks indicate which study was excluded for the $r$-value computation.}\label{fig-CD004421}
\end{figure}

For example, the objective of Review CD004421
was to assess the efficacy of therapy taxane containing chemotherapy regimens
as adjuvant treatment of pre- or post-menopausal women with early breast cancer.
The review included 11 studies, out of which only three studies were significant, and the remaining eight studies had non-significant effects. When combined in a meta-analysis,
the evidence was highly significant, and the review conclusion was
that the use of taxane containing adjuvant chemotherapy regimens  improved the overall survival of women with early breast cancer, see Figure
\ref{fig-CD004421}. 

In order to reverse the significant conclusion, we need to leave out 6 studies: for $u=6$, the  $r$-value was 0.0281, but for $u=7$, the  $r$-value was 0.0628, see Table 3. Therefore, with $95\%$ confidence, the true number of studies with an effect (in the same direction) is at least 6. More generally, testing in order at significance level $\alpha$ , results in a 1-$\alpha$ confidence lower bound on the number of studies with an effect in a fixed-effect meta-analysis (see \cite {Heller11} for proof).
Note that although we have a lower bound on the number of studies that show an effect, we cannot point out to which studies these are. This is so since the pooling of evidence in the same direction in several studies increases the lower bound, even though each study on it's own maybe non-significant.
%(to add a small numeric example in appendix XXX)

\begin{table}[ht]
\caption{The $r$-value excluding $u-1$ studies, for $u=2,\ldots,7$, in Review CD004421. This exclusion is in worst-case order, i.e in order of the study that will show the highest lower bound.}
\centering
\begin{tabular}{rrrrr}
  \hline

&     &  & sensitivity interval                &              \\
& $u$ &  $r$-value             & lower bound & exclued study \\ 
  \hline
  1 & 2 & 2.91e-06 & 0.89 & BCIRG 001 \\ 
  2 & 3 & 1.64e-04 & 0.91 & BIG 2-98 \\ 
  3 & 4 & 2.29e-03 & 0.94 & Taxit 216 \\ 
  4 & 5 & 8.32e-03 & 0.96 & NSABP B-28 \\ 
  5 & 6 & 2.81e-02 & 0.99 & HeCOG \\ 
  6 & 7 & 6.28e-02 & 1.01 & PACS 01 \\ 
  \hline
   
\end{tabular}
\end{table}\label{tab-CD004421}

\subsection{Accounting for multiplicity}\label{subsec-multiplicity}
When more than one primary endpoint is examined, the $r$-value needs to be smaller in order to establish replicability. This is exactly the same logic as with $p$-values, for which we need to lower the significance threshold when faced with multiplicity of endpoints.  Family-wise error rate (FWER) or false discovery rate (FDR) controlling procedures can be applied to the individual $r$-values in order to account for the multiple primary endpoints, see \cite{Benjamini09} for details. 

For example, in Review CD005211 four endpoints were examined, with the following $r$-values: (1) 0.1231; (2) 0.0017; (3) 0.0167; (4) 0.1776. For FWER control over replicability claims at the 0.05 level, the Bonferroni-adjusted $r$-values are the number of endpoints multiplied by the original $r$-values.  Only endpoint (2) is reported as replicated using Bonferroni at the 0.05 level, since it is the only Bonferroni-adjusted $r$-value below 0.05,  $4\times  0.0017<0.05$. 

For FDR control over replicability claims, we can use the Benjamini-Hochberg (BH)  procedure (\cite{Benjamini95}) on the reported $r$-values. The BH-adjusted $r$-values for a sorted list of $M$ $r$-values, $r_{(1)}\leq \ldots \leq r_{(M)}$, are $\min_{i\geq j} \frac{Mr_{(i)}}{i}, \quad  j=1,\ldots, M.$ In Review CD005211, the sorted list is $(0.0017, 0.0167, 0.1231, 0.1776)$ and the adjusted $r$-values are $(0.0068, 0.0334, 0.1641, 0.1776)$. Therefore, endoints (2) and (3), the two endpoints with the smallest $p$-values in the sorted list, are reported as replicated using FDR at the 0.05 level, since for both endpoints the BH-adjusted $r$-values are below 0.05.  

%HOW MANY STUDIES FOR EACH ENDPOINT IN REF32? ARRAY OF P-VALUES PER ENDPOINT? 

%HOW MANY PRIMARY ENDPOINTS IN OUR TWO DOMAINS? SAY SOMETHING ABOUT CORRECTING FOR THE MULTIPLICITY IN THE NUMBER OF PRIMARY OUTCOMES? In a typical
%systematic review, although there is usually one (or two) primary
%endpoints,
% many additional endpoints are analyzed. Some, but not all, of the endpoints are expected to have replicated findings.
% For a full replicability analysis of all the endpoints, while controlling for false positives, it is necessary to use more advanced methods,
% along the lines suggested in \cite{Bogomolov13}, \cite{Heller14b}, and Bogomolov and Heller arXiv (TBA).
% These methods were designed for replicability analysis when two studies are available, and to use them will entail computing two meta-analysis $p$-values, or examine all the pairs of studies, on all the endpoints XXX.  This problem will be dealt with in a future paper. 

\section{Discussion}
In this work we suggested enhancing the systematic reviews meta-analyses, for both fixed effect and random effects model, with a measure that quantifies the strength of replicability,i.e., the $r$-value. In the reporting, if the $r$-value is small we have evidence that the conclusion is based on more than one study, i.e., that the effect was replicated across studies. We suggest adding a cautionary note if the $r$-value is greater than the significance level (say $=0.05$), that states that the conclusion
depends critically on a single study. This does not mean that the
conclusion is necessarily reversed, but the large $r$-value warrants
another examination of the studies in the meta-analysis, and if the
single study upon which the review relies was very well conducted
the conclusion may still be justified despite it being only a single
study.

We would like to emphasize that replicability analysis is relevant for both fixed effect and random effects model meta analysis. In both cases, the meta-analysis can be significant even though the true summary effect is  greater than zero in only one study out of the $N$ and hence the replicability analysis is needed.  Specifically, for the random effect model in Appendix \ref{app-sim} we show simulations where $N-1$ studies have effects $\theta_i$ samples for the normal distribution with zero mean, and one study has effect $\mu_n \in \{0,\ldots, 5\}$. When $\mu_n=0$, the fraction of times the null is rejected at the nominal 0.05 level using a $t$-test with $N-1$ degrees of freedoms on the   sample of $N$ estimated effect sizes is about 0.05, and using the meta-analysis computations suggested in \cite{Higgins11} the fraction is at most 0.12. However, when $\mu_n>0$,  the fraction of times the null is rejected at the nominal 0.05 level using a $t$-test with $N-1$ degrees of freedoms on the sample of $N$ estimated effect sizes can be as high as 0.15, and using the meta-analysis computations suggested in \cite{Higgins11} the fraction  can reach almost 0.3. We conclude from these simulations that for meta-analysis, it is better to use the $t$-test, and that even with this non-liberal test the significant conclusion can be entirely driven from a single study. Therefore, a replicability analysis is necessary in order to rule out the possibility that a significant random effect meta-analysis conclusion is driven by a single study. 
%the common usage of computing significance using the normal distribution when calculating the random effect meta-analysis p-value (instead of using the $t$-distribution with $N-1$ degrees of freedom) results in a type I error rate substantially greater than 5\% under the null hypothesis

In our two domains there were typically 1-4 primary endpoints per review. We briefly discussed ways to account for the multiplicity of primary endpoints in assessing replicability in Section \ref{subsec-multiplicity}. We regard this as an extension since the emphasis, and the new contribution, of this paper is the introduction of the 
 $r$-value into the meta-analysis conclusions.%, and adjusting for multiplicity is of secondary importance. 

%WE DO NOT DISCUSS NONSIG TO SIG UPON SINGLE STUDY REMOVAL: WE SHOULD
%ADDRESS THIS FOR ADVERSE EVENTS? 
%We suggest computing the $r$-value only to endpoints for which the meta-analysis result is significant. Although it is possible that a non-significant meta-analysis result will become significant upon exclusion of a single study, we believe that such an analysis can result in undesired, anti-conservative results. Specifically, if an investigators has no finding, but then upon looking at a subset a finding is discovered, and then reported as a finding, then the reporting must take into account the multiplicity of all subsets in order to be valid. 

%DISCUSSION OF OTHER COMBINING METHODS FOR FIXED EFFECTS? FOR RANDOM EFECTS? 
%For fixed effects, can combine the $p$-values instead of the fixed
%effect meta-analysis combining method, as detailed in \cite{conj}.
%SHOW AN EXAMPLE WHEN IT MATTERS?

%IN THIS WORK THE ANALYSIS DOES NOT POINT OUT TO THE SINGLE STUDY
%THAT REVERSES THE CONCLUSION: IF WE WANT TO ACTUALLY POINT OUT WE
%NEED TO CONTROL FOR MULTIPLE HYPOTHESES. DO WE WANT TO SAY SOMETHING
%ABOUT THIS?

%DISCUSSION OF R-VALUE VERSUS HETEROGENEITY IN RANDOM EFFECTS?
%FUTURE WORK: CONNECTION BETWEEN THE FIXED AND RANDOM EFFECT R-VALUES?

%\bibliographystyle{apalike}
%\bibliography{references}

\appendix
\section{Examples for the influenza domain}\label{app-influenza}
\begin{figure}[ht]
\centering
\includegraphics[width=0.3\textwidth, height=0.3\textheight]{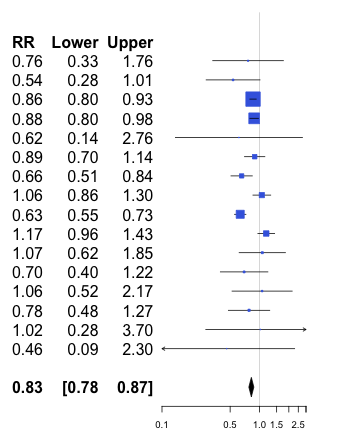}
\includegraphics[width=0.3\textwidth, height=0.3\textheight]{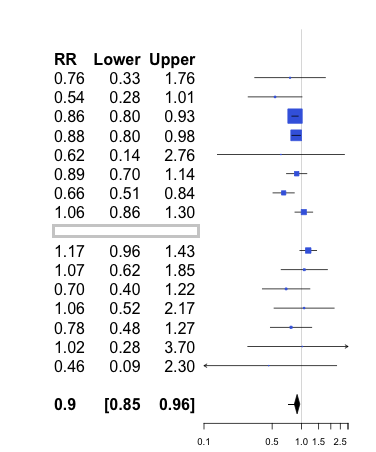}
\caption{The forest plot in Review CD001269 (Left) and excluding study 9 (Right). The $r$-value was  0.0014. The sensitivity interval, [0.75,0.96],  is the confidence interval excluding study 9 (the black diamond in the right panel) with an additional (very small) left tail. The axis is on the logarithmic scale.}\label{fig-Influenza-like illness}
\end{figure}

Our first example is based on a meta-analysis in review CD001269, analyzed by the authors as a random effect meta-analysis, which is non-sensitive and thus has a small $r$-value. The objective of Review  CD001269 
was to assess the effects of vaccines against influenza in healthy adult. 
Four studies were significant (all favoured treatment), and the remaining 
twelve studies had non-significant effects at the .05 significance level (seven favoured the treatment, five favoured the control).
When combined in a meta-analysis the evidence was significant, and the review conclusion was
that the placebo arm had a higher risk of Influenza-like illness than the vaccine arm, see the left panel of Figure
\ref{fig-Influenza-like illness}. Even after removing each significant study (in particular
the most influential: study number 9) there was still a significant
effect in the meta-analysis at the 0.05 level, see the right panel of Figure
\ref{fig-Influenza-like illness}. The $r$-value is 0.0014. The significant meta-analytic
conclusion can therefore be accompanied by a statement that the
replicability claim is established at the .05 level of significance.
This is a stronger scientific claim than that of the meta-analysis, 
and it is supported by the data in this example.  

\begin{figure}[ht]
\centering
\includegraphics[width=1\textwidth, height=0.6\textheight]{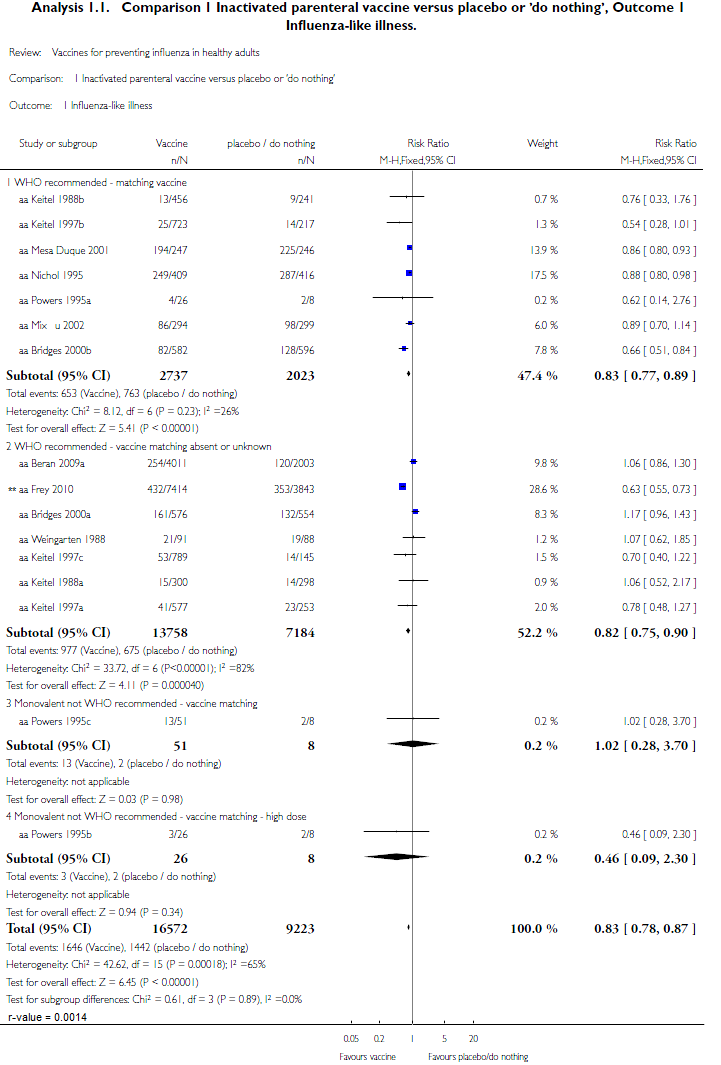}
\caption{The forest plot in the original Review CD001269 including the $r$-value, which  was  0.0014. The asterisks indicate which study was excluded for the $r$-value computation.}\label{fig-Influenza-like illness-orig}
\end{figure}

We suggest accompanying the original forest plot
with this $r$-value, see Figure \ref{fig-Influenza-like illness-orig}.
Note that although the referred meta-analysis is significant and the replicability claim is established at the .05 level of significance, in the main results of Review CD001269 the authors write :
"The overall effectiveness of parenteral inactivated vaccine against influenza-like illness (ILI) is limited, corresponding to a number needed to vaccinate (NNV) of 40 (95\% confidence interval (CI) 26 to 128)". 
to this, we suggest adding the following information: ``This result was replicated in more than one study ($r$-value = 0.0014)". The replicability claim is relevant even in the presence of a limited effect size: at least two studies showed that there is a (possibly limited)
effect of parenteral inactivated vaccine against influenza-like illness (ILI) is limited.

%XXX is the smallest LB 

\begin{figure}[ht]
\centering
\includegraphics[width=0.4\textwidth, height=0.3\textheight]{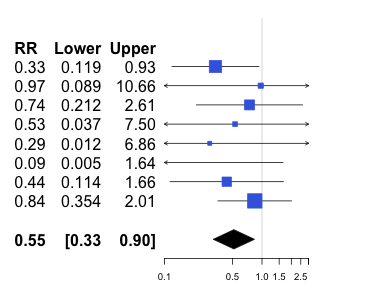}
\includegraphics[width=0.4\textwidth, height=0.3\textheight]{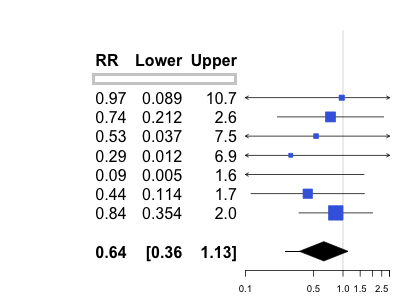}
\caption{The forest plot in Review CD008965 (Left) and excluding study 1 (Right). The $r$-value was 0.1206. The sensitivity interval, [0.24, 1.13], is the confidence interval excluding study 1 (the black diamond in the right panel) with an additional (small) right tail. The axis is on the logarithmic scale.}\label{fig-Neuraminidase inhibitors}
\end{figure}

Our second example, also from the influenza domain, is based on a meta-analysis in review CD001269, analyzed by the authors as a random effects meta-analysis. In this example the random effect meta-analysis was sensitive. The objective of 
Review CD008965 was to describe the potential benefits and harms of
Neuraminidase inhibitors for influenza in all age groups. In the meta-analysis a significant
finding was discovered, see the left panel of Figure \ref{fig-Neuraminidase inhibitors}. 
Note that only one study was significant and the remaining seven studies were not significant 
(with large confidence intervals). 
After removing study number 1,there was no longer a significant effect in the meta-analysis, see
the right panel of Figure \ref{fig-Neuraminidase inhibitors}. 
The $r$-value was 0.1206 based on the random effect meta-analysis computations as suggested in \cite{Higgins11}, and 0.0661 based on the meta-analysis computations as suggested in \cite{IntHout14}. The replicability claim was not established at the .05 level
of significance. This lack of replicability, quantified by the $r$-value, cautions practitioners
that the significant meta-analysis finding may depend critically on
a single study. 

\begin{figure}[ht]
\centering
\includegraphics[width=1\textwidth, height=0.5\textheight]{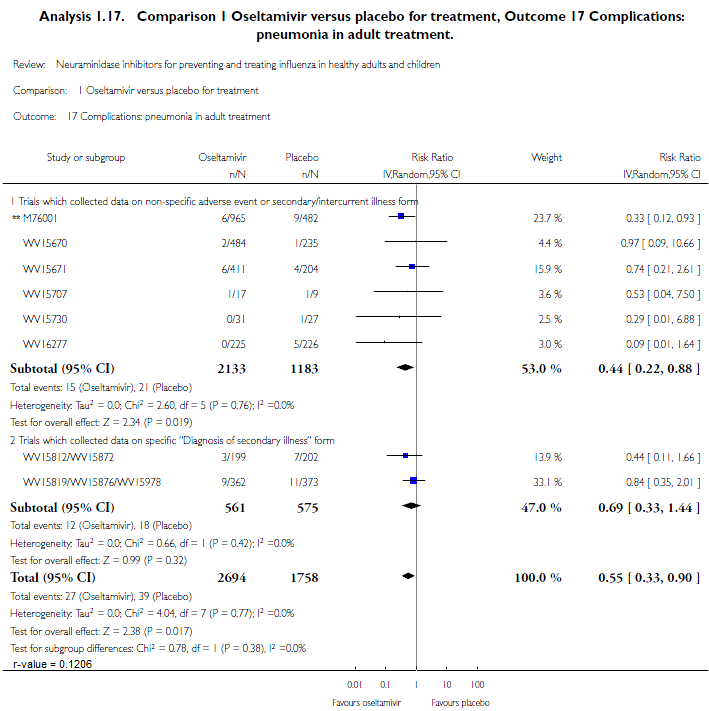}
\caption{The forest plot in the original Review CD008965 including the $r$-value, which  was  0.1206. The asterisks indicate which study was excluded for the $r$-value computation.}\label{fig-Neuraminidase inhibitors-orig}
\end{figure}

We suggest accompanying the original forest plot
with this $r$-value, see Figure \ref{fig-Neuraminidase inhibitors-orig}.
In the main results of Review CD008965 the authors write "In adults treatment trials, Oseltamivir significantly reduced self reported, investigator-mediated, unverified pneumonia (RR 0.55, 95\% CI 0.33 to 0.9)" ; 
To this, we suggest adding the following information: ``We cannot rule out the possibility that this result is based on a single study ($r$-value = 0.1206)". %XXX is the largest UB
Note that the conclusion was not that complication were reduced, but this was due to lack of diagnostic definitions.
The authors’ conclusion in this review was that "treatment trials with
oseltamivir do not settle the question of whether the complications of influenza (such as pneumonia) are reduced, because of a lack of diagnostic definitions". 

%Note: this result was not replicated in more than one study (r-value = 0.24).

\section{Sensitivity analysis computation details}\label{app-computations}
Let $p^L_{i_1,\ldots,i_k}$ and $p^R_{i_1,\ldots,i_k}$ be, respectively, the left- and right- $p$-values from a meta-analysis on the subset $(i_1,\ldots,i_k)\subset \{1,\ldots,N\}$ of the $N$ studies in the full meta-analysis, $k<N$. Let $\Pi(k)$ denote the set of all possible subsets of size $k$. 

\subsection{The $r$-value computation}\label{app-subsec-rvalue}
For a meta-analysis based on $N$ studies, a replicability claim is a claim that the conclusion remains significant (e.g., rejection of the null hypothesis of no treatment effect) using a meta-analysis of each of the $\binom{N}{u-1}$ subsets of $N-u+1$ studies, where $u=2,\ldots, N$ is a parameter chosen by the investigator. 
Specifically, for $u=2$, a replicability claim is a claim that the conclusion remains significant  using a meta-analysis of each of the $N$ subsets of $N-1$ studies.

The $r$-value for replicability analysis, where we claim replicability if the conclusion remains significant using a meta-analysis of each of the $\binom{N}{u-1}$ subsets of $N-u+1$ studies is computed as follows. For left- sided alternative, the $r$-value is 
$$r^L = \max_{(i_1,\ldots,i_{N-u+1})\in \Pi(N-u+1)} p^L_{i_1,\ldots,i_{N-u+1}}.$$
For right- sided alternative, the $r$-value is 
$$r^R = \max_{(i_1,\ldots,i_{N-u+1})\in \Pi(N-u+1)} p^R_{i_1,\ldots,i_{N-u+1}}.$$
For two-sided alternavies, the $r$-value is 
$$r = 2\min(r^L,r^R).$$ 

%More formally, 

\subsection{Sensitivity analysis for confidence intervals}
The sensitivity interval is the union of all the meta-analysis confidence intervals using the  
$\binom{N}{u-1}$ subsets of $N-u+1$ studies.   
 The upper limit of the $(1-\alpha)$ sensitivity interval is the upper limit of the $(1-\alpha)$ confidence interval from the meta-analysis on $(i_1^L, \ldots, i^L_{N-u+1})$, where  $(i_1^L, \ldots, i^L_{N-u+1})$ is the  subset that achieves the maximum $p$-value for the left-sided $r$-value computation.  Similarly, the lower limit of the $(1-\alpha)$ sensitivity interval is the lower limit of the $(1-\alpha)$ confidence interval from the meta-analysis on $(i_1^R, \ldots, i^R_{N-u+1})$, where  $(i_1^R, \ldots, i^R_{N-u+1})$ is the  subset that achieves the maximum $p$-value for the right-sided $r$-value computation.

The meta-analysis is non-sensitive (at the desired value of $u$) if and only if the sensitivity interval does not contain the null hypothesis value. This follows from the following argument. 
 To see this, note that  $r\leq \alpha$, if and only if  $r^L\leq \alpha/2$ or $r^R\leq \alpha/2$. Since $r^L\leq \alpha/2$ if and only if the upper limit of all the meta-analysis $1-\alpha$ confidence  intervals of subsets of size $N-u+1$ is below the null value, and $r^R\leq \alpha/2$ if and only if the lower limit of all the meta-analysis $1-\alpha$ confidence  intervals of subsets of size $N-u+1$ is above the null value, the result follows.

 \subsection{Leave-one-out sensitivity procedure}

For meta-analysis with N studies and significant effect size $\theta < \theta_{0}$ ,where $\theta_{0}$ is the null effect ,e.g., 1 for HR  (two- sided alternative):
\newline
1) Compute meta-analysis of each of the $\binom{N}{u-1}$ subsets of $N-u+1$ studies.
\newline
2) Choose the $N-u+1$ subset of studies that achieves the maximum $p$-value for the left-sided $r$-value computation:
$(i_1^L, \ldots, i^L_{N-u+1})$.
\newline
3) Compute the two-sided $r$-value : $$r = 2\min(r^L,r^R).$$
4) If the $r$-value $\leq$ 0.05, the replicability is established in at lease $u$ studies. Otherwise, the replicability
is established in at most $u-1$ studies (for $u$=2 ,  $r$-value $>$ 0.05 means that the finding is not replicable).

For meta-analysis with N studies and significant effect size $\theta > \theta_{0}$ ,where $\theta_{0}$ is the null effect ,e.g., 1 for HR  (two- sided alternative):
\newline
1) Compute meta-analysis of each of the $\binom{N}{u-1}$ subsets of $N-u+1$ studies.
\newline
2) Choose the $N-u+1$ subset of studies that achieves the maximum $p$-value for the right-sided $r$-value computation:
$(i_1^R, \ldots, i^R_{N-u+1})$. 
\newline
3) Compute the two-sided $r$-value : $$r = 2\min(r^L,r^R).$$
4)	If the $r$-value $\leq$ 0.05, the replicability is established in at lease $u$ studies. Otherwise, the replicability
is established in at most $u-1$ studies (for $u$=2 ,  $r$-value $>$ 0.05 means that the finding is not replicable).

 \section{Random effects meta analysis simulation}\label{app-sim}
Using the following simulation we demonstrate that a significant random effect meta-analysis is not equivalent to replicability. %testing for replicability.
Meaning, random effect meta-analysis can be significant even though the effect is greater than zero in only one study out of $N$. We show that the probability of rejecting the null hypothesis with a single outlying study can be as high as 6 times the nominal level using the 
meta-analysis computations of \cite{Higgins11}, and as high as 3 times the nominal level using the more conservative approach of \cite{IntHout14}.

For $N \in \{3,5,7,9,20 \}$ studies, we sampled 
 $N-1$   effects $\mu_{i}, i=1,..,N-1 $  from the distribution 
$N(0,\tau^2)$, where $\tau^2\in \{0.01,0.04, 0.09,0.25,0.49,1\}$. For the $N$th study, the effect was $\mu_n \in \{0, 0.05, 0.1, \ldots ,4.5 ,5\}$. For each study $i \in \{1,\ldots,N\}$, we sampled 
 observed effects $ \hat \theta_i$ from the normal distribution with mean $\mu_i$ and standard deviation 0.01. We computed the random effect meta-analysis one-sided p-value using the computations suggested in \cite{Higgins11}, i.e., using the $z$-test on the average observed effects, as well as using the $t$-test on the sample of observed effects. 
We estimated the probability of rejection the null hypothesis of zero mean based on  $10^4$ iterations.

\begin{figure}[ht]
\centering
\includegraphics[width=0.3\textwidth, height=0.35\textheight]{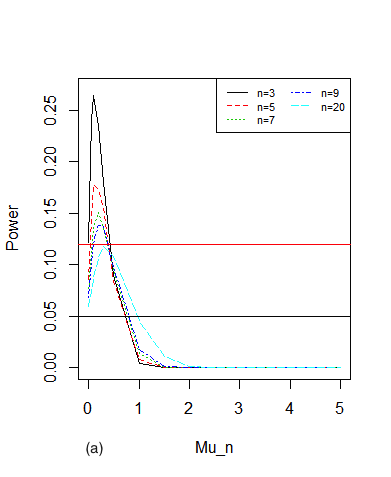}
\includegraphics[width=0.3\textwidth, height=0.35\textheight]{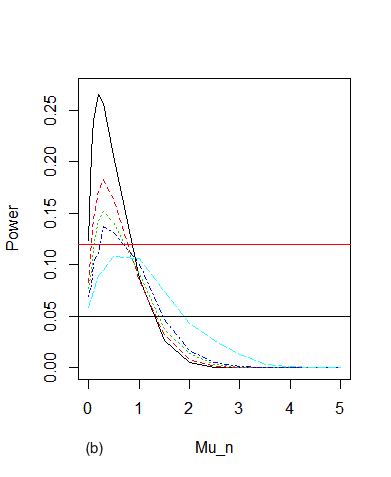}
\includegraphics[width=0.3\textwidth, height=0.35\textheight]{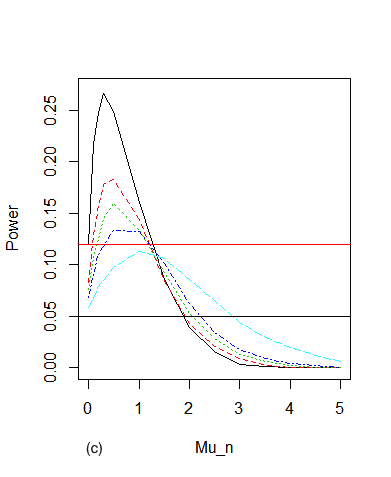}
\includegraphics[width=0.3\textwidth, height=0.35\textheight]{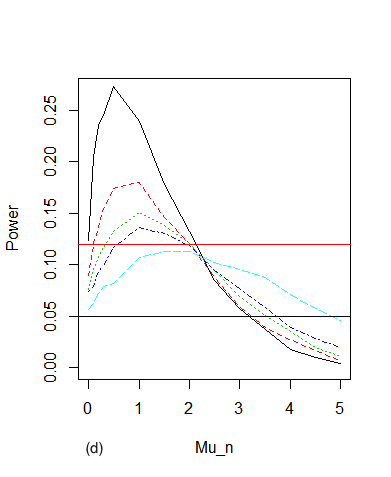}
\includegraphics[width=0.3\textwidth, height=0.35\textheight]{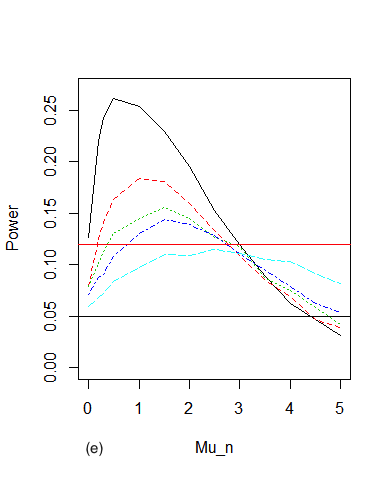}
\includegraphics[width=0.3\textwidth, height=0.35\textheight]{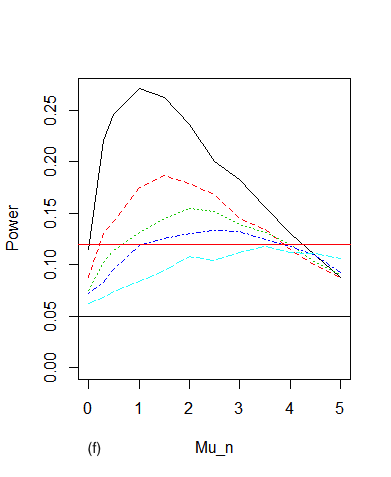}
\caption{The average fraction of rejections at the 0.05 level, using the $z$-test detailed in page 74 of \cite{Higgins11}: (a) $\tau^2 =  0.01$; (b) $\tau^2 =  0.04$; (c) $\tau^2 =  0.09$; (d) $\tau^2 =  0.25$; (e) $\tau^2 =  0.49$; (f) $\tau^2 =  1$.  The probability of the type I error is above $0.05$ for all $N$s and $\tau$s, ranging from about $\leq 12\%$ for $N=3$ and decreasing to $0.055$ for $N=20$. The maximum fraction of rejections is $0.26$ for $N=3$ and $\tau^2=0.01$, and decreases for increasing $N$ and $\tau$. The range of values of $\mu_N$ for which it is above 5\% increases with $\tau^2$ and with $N$. Even if the $z$-test is acceptable for meta-analysis when the number of studies is large enough, we still have a problem with lack of replicability.}\label{fig-ztest}
\end{figure}

\begin{figure}[ht]
\centering
\includegraphics[width=0.3\textwidth, height=0.35\textheight]{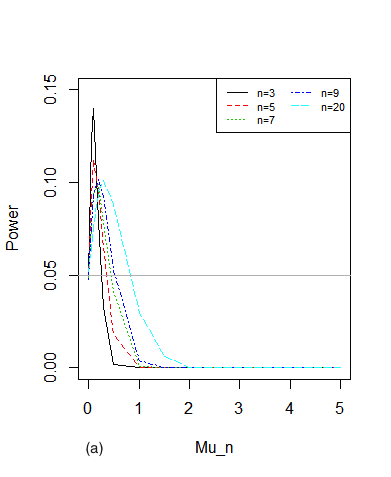}
\includegraphics[width=0.3\textwidth, height=0.35\textheight]{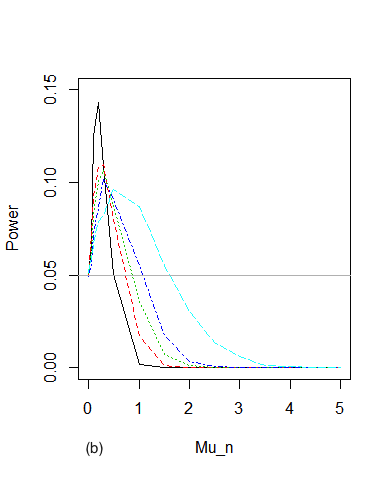}
\includegraphics[width=0.3\textwidth, height=0.35\textheight]{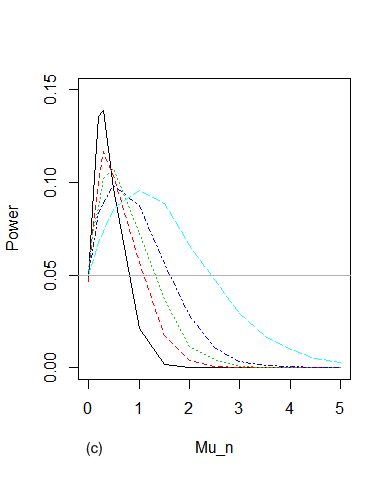}
\includegraphics[width=0.3\textwidth, height=0.35\textheight]{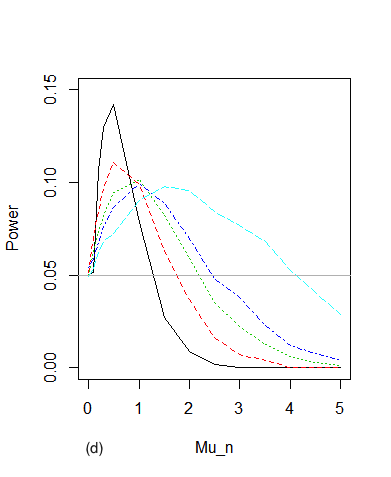}
\includegraphics[width=0.3\textwidth, height=0.35\textheight]{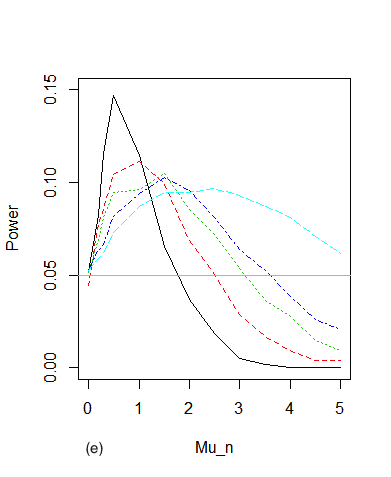}
\includegraphics[width=0.3\textwidth, height=0.35\textheight]{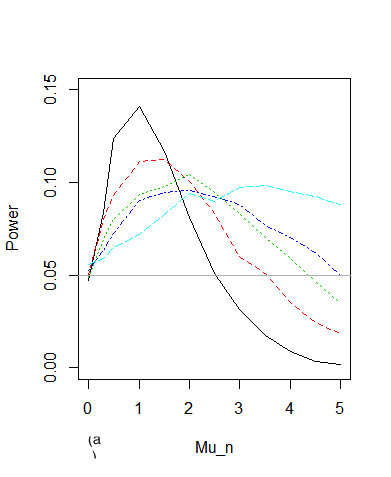}
\caption{The average fraction of rejections at the 0.05 level, using the $t$-test: (a) $\tau^2 =  0.01$; (b) $\tau^2 =  0.04$; (c) $\tau^2 =  0.09$; (d) $\tau^2 =  0.25$; (e) $\tau^2 =  0.49$; (f) $\tau^2 =  1$.  The probability of the type I error is $0.05$ for all $N$s and $\tau$s. The maximum fraction of rejections is $0.15$ for $N=3$ and $\tau^2=0.01$, and decreases for increasing $N$ and $\tau$. The range of values of $\mu_N$ for which it is above 5\% increases with $\tau^2$ as well as with $N$. Even though the $t$-test controls the probability of type I error for meta-analysis, we still have a problem with lack of replicability.}\label{fig-ttest}
\end{figure}

Figures \ref{fig-ztest} and \ref{fig-ttest} show the resulting estimated probability of rejecting the null hypothesis.  The random effect meta-analysis is significant in more than 
5\% of the iterations for all $N$ in values of $\mu_N>0$ that are not too large relative to the value of $\tau$.  
The larger the value of $\tau^2$, the greater the range of $\mu_N>0$ for which the nominal level of significance is not maintained.

In \cite{Higgins11}, the normal distribution is used for the random effect meta-analysis $p$-value, instead of the $t$-distribution with $N-1$ degrees of freedom which in our simulation (with equal study weights) results in an exact $\alpha=0.05$ level test when $\mu_N=0$. We see that the usage of the $z$-test instead of the $t$-test results in a type I error rate substantially greater than 5\% under the null hypothesis (i.e. $\mu_N = 0 $) and in a higher rejection rate of the null hypothesis for $\mu_N > 0 $ in comparison to the fraction of rejections using the $t$-test when there is no replicability.
 
%null hypothesis rejection rate

\end{document}